\documentclass{article}

\usepackage{spconf}

\usepackage{amsmath,amssymb,amsfonts}
\usepackage{graphicx}
\usepackage{booktabs}
\usepackage{xcolor}
\usepackage[hidelinks]{hyperref}

\title{ADAPTING SLEEP FOUNDATION MODELS FOR DENSE SLEEP SPINDLE LOCALIZATION}

\name{
Alexis Bollengier$^{\dagger \star}$,
Abel Díaz Berenguer$^{\dagger}$,
Olivier Pallanca$^{\star}$,
André Anjos$^{\star}$ \&
Hichem Sahli$^{\dagger \ddagger}$ 
}

\address{
$^{\star}$ Idiap Research Institute, Martigny, Switzerland\\
$^{\dagger}$ Department of Electronics and Informatics (ETRO), Vrije Universiteit Brussel (VUB), Brussels, Belgium\\
$^{\ddagger}$ Interuniversity Microelectronics Centre (IMEC), Heverlee, Belgium
}

\begin{document}

\maketitle


\begin{abstract}
Adapting sleep foundation models to dense electroencephalography (EEG)
event localization is challenging because multi-second representations may
not preserve sufficient temporal detail for precise boundary recovery.
We address sleep spindle localization by adapting a frozen SleepFM
encoder with a local time-frequency pathway that preserves spindle-relevant
spectral content and intra-token position. The local representation drives
localization, while residual cross-attention incorporates SleepFM temporal
context. Dense predictions are obtained through localized token-to-sample
reconstruction and temporal refinement. Tested on the Massive Online Data Annotation (MODA), the proposed model achieves
an event-level F1 of 0.8080$\pm$0.0040 at IoU $0.2$ across six folds.
Ablations show that localization primarily depends on local spectral and
sub-token temporal information, while the pretrained representation provides
complementary context. These results indicate that dense adaptation benefits
from preserving information below the pretrained token scale. Code will be released upon acceptance.
\end{abstract}



\vspace{-5mm}
\section{Introduction}
Sleep spindles are transient EEG events in non-rapid eye movement sleep,
typically lasting 0.5--3 s with sigma-band activity around 11--16 Hz
\cite{fernandez2020sleep}. Their characteristics are linked to memory
consolidation and investigated as biomarkers for neuropsychiatric disorders
\cite{manoach2019spindles}, motivating accurate detection and boundary
delineation.

Dense detectors such as SUMO \cite{kaulen2022sumo} and BLAST
\cite{guo2025blast} learn task-specific spindle representations, whereas
sleep foundation models learn reusable representations from large
polysomnography datasets. SleepFM \cite{thapa2025sleepfm500k} supports
multiple downstream tasks but compresses multi-second EEG windows into
compact tokens, creating a resolution mismatch for dense localization.
SleepGPT \cite{huang2026sleepgpt} supports spindle detection by jointly
learning temporal and frequency-domain representations during pretraining.
Here, rather than pretraining a new time-frequency backbone, we adapt a
frozen representation with task-specific local information to recover
boundaries below the pretrained token scale.

The proposed architecture keeps a pretrained SleepFM encoder frozen and adds a local time-frequency branch that preserves spindle-related
spectral information and intra-token position. Residual cross-attention
incorporates the pretrained SleepFM representation as complementary context, followed by sample-level reconstruction of sub-token temporal structure.

The main contributions are: (1) dense sample-level adaptation of a frozen SleepFM encoder; (2) a physiologically structured time-frequency
representation preserving local spectral content and intra-token position, with asymmetric cross-attention to the pretrained SleepFM representation; and (3) localized token-to-sample reconstruction with reinjection of retained intra-token features. The approach is evaluated on the MODA benchmark using MASS EEG segments and expert-consensus spindle annotations across six published folds, with controlled architectural ablations.

\begin{figure*}[t]
    \centering
    \includegraphics[width=\textwidth, trim=0cm 0.3cm 0cm 0.7cm,
    clip]{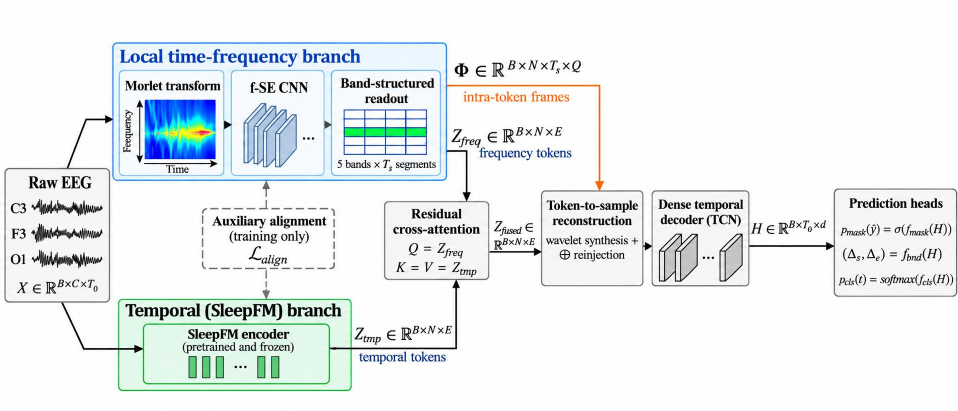}
    \caption{Overview of the proposed spindle localization architecture. The frozen SleepFM encoder provides temporal context, while a local time-frequency branch preserves spectral and intra-token information for dense reconstruction.}
    \label{fig:architecture}
\end{figure*}
\section{Methods}
\subsection{Overview}
The proposed architecture addresses the temporal-resolution mismatch between pretrained sleep representations and dense spindle localization. When local structure is compressed into a multi-second token, increasing the output resolution alone cannot recover the precise location of spindle-related activity. This adaptation therefore preserves local information in a parallel branch and reuses it during sample-level reconstruction.

Let $\mathbf{X}\in\mathbb{R}^{B\times C\times T_0}$ represent a multichannel input segment. We use a SleepFM encoder pretrained under the original multimodal contrastive formulation \cite{thapa2025sleepfm500k}. 
Its channel-agnostic design accommodates heterogeneous recording configurations and modality-specific representations, motivating its use with a reduced configuration of brain activity signals (BAS) for downstream spindle localization.
The pretrained weights are kept frozen, while the input window duration is adapted from 5 to 3 seconds. For the downstream spindle-localization task, the encoder processes the available EEG/EOG channels to produce temporal tokens
$\mathbf{Z}_{\mathrm{tmp}}\in\mathbb{R}^{B\times N\times E}$,
where $E=128$. A task-specific time-frequency branch operates on the same grid, generating frequency tokens $\mathbf{Z}_{\mathrm{freq}}$ and ordered intra-token features $\mathbf{\Phi}$. Figure~\ref{fig:architecture} summarizes their integration before sample-level prediction.

\subsection{Local Time-Frequency Representation}

\begin{figure}[t]
    \centering
    \includegraphics[width=\columnwidth]{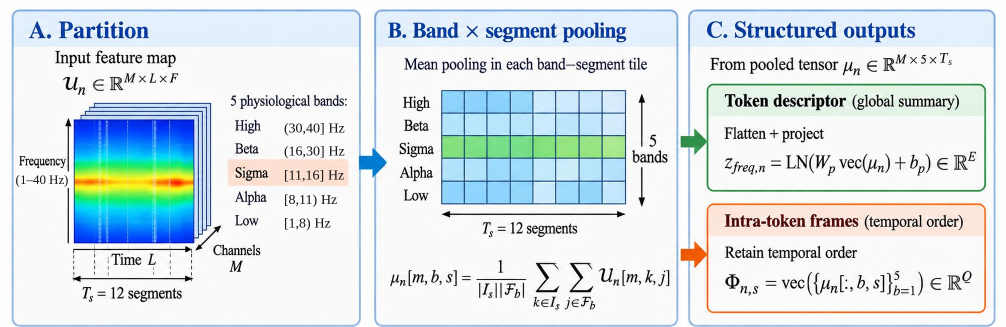}
    \caption{Band-structured readout. Mean pooling across physiological bands and ordered temporal segments produces a token descriptor and intra-token features.}
    \label{fig:band_readout}
\end{figure}

The local time-frequency branch captures sigma-band oscillations and the transient morphology of sleep spindles. A fixed Morlet wavelet transform represents the EEG jointly in time and frequency \cite{cohen2019morlet}. The resulting scalogram is segmented using the same overlapping windows as SleepFM. Each patch is processed by a frequency-wise Squeeze-and-Excitation module adapted from \cite{hu2018squeeze}, emphasizing informative frequency regions before three $3\times3$ convolutional layers with $M=64$ output features extract local patterns.

A key contribution is the band-structured readout in Fig.~\ref{fig:band_readout}. Global pooling retains the presence of relevant activity but discards its position within the 3-second token. Instead, the representation is organized into five physiological frequency bands and $T_s=12$ ordered temporal segments. Mean pooling over the scalogram-frame indices $\mathcal{I}_s$ of temporal segment $s$ and the frequency indices $\mathcal{F}_b$ of band $b$ preserves spectral organization and intra-token position. The resulting representation forms $\mathbf{Z}_{\mathrm{freq}}$ for token-level fusion, while the same features are retained as $\mathbf{\Phi}$ for dense reconstruction, with $Q=5M=320$ features per temporal segment.

\subsection{Context Fusion}

The frequency representation preserves localized spindle evidence, whereas SleepFM provides broader pretrained temporal context. Their interaction is therefore asymmetric. Frequency tokens query the pretrained sequence using residual multi-head cross-attention \cite{vaswani2017attention}
\[
\mathbf{Z}_{\mathrm{attn}}
=
\mathrm{LN}\!\left(
\mathbf{Z}_{\mathrm{freq}}
+
\mathrm{CrossAttn}
\left(
\mathbf{Z}_{\mathrm{freq}},
\mathbf{Z}_{\mathrm{tmp}},
\mathbf{Z}_{\mathrm{tmp}}
\right)
\right)
\]
A residual feed-forward sublayer then produces $\mathbf{Z}_{\mathrm{fused}}$. This keeps $\mathbf{Z}_{\mathrm{freq}}$ as the main localization pathway while retrieving complementary context from SleepFM.

\subsection{Localized Token-to-Sample Reconstruction}

Fusion still operates at 3-second token resolution. The decoder must therefore reconstruct temporal structure within each token rather than simply increase the sequence length. Localized wavelet synthesis lets each fused token predict both the temporal patterns contributing to reconstruction and their position within the window.

Let $\mathbf{z}_n$ denote fused token $n$ and $W=384$ its window length. A fixed bank $\mathbf{\Psi}\in\mathbb{R}^{K_a\times P_a\times W}$ contains $K_a=16$ temporal atom types at $P_a=6$ candidate positions. The atoms span one to four cycles with Gaussian widths logarithmically spaced between $W/16$ and $W/2.5$, with candidate centers from $0.20W$ to $0.80W$. They describe event-scale temporal variations rather than sigma oscillations, which are already represented by the frequency branch.

Each token predicts feature coefficients $\mathbf{C}_n\in\mathbb{R}^{K_a\times d}$ with $d=64$ and position probabilities $\pi_{n,k,p}$. The localized form of atom $k$ is
\[
A_{n,k,w}
=
\sum_{p=1}^{P_a}
\pi_{n,k,p}\Psi_{k,p,w}
\]
The sample-wise feature reconstruction is
\[
\mathbf{X}_{n,w}
=
\mathrm{LN}
\left(
\mathbf{W}_b\mathbf{z}_n+\mathbf{b}_b
\right)
+
\sum_{k=1}^{K_a}
A_{n,k,w}\mathbf{C}_{n,k}
\]
The atom bank remains fixed, while its coefficients and placement probabilities are predicted from each token. This constrains reconstruction to localized temporal patterns rather than an unconstrained sample-level sequence.

Because consecutive SleepFM windows overlap, their reconstructions are combined using Hann-weighted overlap-add 
\[
\mathbf{Z}_{\mathrm{sample}}[t]
=
\frac{
\sum_{(n,w)\in\mathcal{I}_t}
g_w\mathbf{X}_{n,w}
}{
\sum_{(n,w)\in\mathcal{I}_t}g_w
}
\]
where $\mathcal{I}_t=\{(n,w)\mid a_n+w=t\}$, $a_n$ is the start of token $n$ and $g_w$ is the Hann weight.

For the reconstruction ablation in Table~\ref{tab:ablations},
localized wavelet synthesis was replaced by linear interpolation of the projected fused token representations between consecutive token centers onto the sample grid, while all other components were kept identical to the frame-off reference.

\subsection{Intra-Token Reinjection and Temporal Refinement}

Wavelet synthesis reconstructs temporal structure from $\mathbf{Z}_{\mathrm{fused}}$, while $\mathbf{\Phi}$ retains explicit intra-token spectral position before compression. These features are projected to the decoder width and expanded from $T_s$ temporal segments to sample positions using nearest-neighbor upsampling. Overlapping contributions are averaged and concatenated with $\mathbf{Z}_{\mathrm{sample}}$, preserving information that would otherwise need to be recovered from the fused token.

A Temporal Convolutional Network \cite{bai2018tcn} with five residual blocks, kernel size 7 and dilations $1,2,3,4,6$, adds neighboring temporal context without reducing sample resolution. Three prediction heads estimate spindle probability, onset and offset distances and the background, boundary or spindle-interior class at each sample.

\subsection{Training Objective}

Training jointly supervises spindle presence, event boundaries and temporal structure. The segmentation objective combines focal loss \cite{lin2017focal} and Dice loss \cite{milletari2016vnet}
\[
\mathcal{L}_{\mathrm{seg}}
=
\mathcal{L}_{\mathrm{focal}}
+
\mathcal{L}_{\mathrm{dice}}
\]
Boundary regression is restricted to annotated spindle samples and optimized using Smooth L1 loss \cite{girshick2015fastrcnn} on onset and offset distances normalized by the sampling frequency. The temporal-classification head applies weighted cross-entropy over background, boundary and spindle-interior samples.

An auxiliary objective regularizes the task-specific frequency representation toward the pretrained SleepFM representation without modifying the frozen backbone. Corresponding tokens are represented as diagonal Gaussian distributions in a shared latent space \cite{kingma2014vae}, with the temporal representation serving as the fixed reference
\[
\mathcal{L}_{\mathrm{cons}}
=
\frac{1}{BN}
\sum_{i=1}^{B}
\sum_{n=1}^{N}
D_{\mathrm{KL}}
\left(
q_{\mathrm{tmp}}^{(i,n)}
\|
q_{\mathrm{freq}}^{(i,n)}
\right)
\]
A complementary frequency constraint encourages spindle-related spectral structure. For each event $e$, sigma prominence compares energy in $[11,16]$ Hz with the surrounding context $[5,11)\cup(16,25]$ Hz
\[
\rho_e
=
\frac{E_{\sigma,e}}
{E_{\sigma,e}+E_{\mathrm{ctx},e}}
\]
The frequency loss averages the prominence deficit over the set $\mathcal{E}$ of predicted event candidates anchored at positive-scoring time-frequency frames with valid sample support,

\[
\mathcal{L}_{\mathrm{freq}}
=
\frac{1}{|\mathcal{E}|}
\sum_{e\in\mathcal{E}}
\max(0,0.70-\rho_e)
\]
The complete objective is
\[
\mathcal{L}
=
\lambda_{\mathrm{seg}}\mathcal{L}_{\mathrm{seg}}
+
\lambda_{\mathrm{bnd}}\mathcal{L}_{\mathrm{bnd}}
+
\lambda_{\mathrm{cls}}\mathcal{L}_{\mathrm{cls}}
+
\lambda_{\mathrm{align}}
\left(
\mathcal{L}_{\mathrm{cons}}
+
\mathcal{L}_{\mathrm{freq}}
\right)
\]

\begin{table*}[t]
\centering
\caption{Controlled architectural ablations on MODA (mean $\pm$ SD across
six training/validation folds on the fixed test set). The frame-off reference
uses frequency-query fusion, both representations, structured readout and
localized wavelet reconstruction; each other configuration changes one
component.}
\label{tab:ablations}
\small
\setlength{\tabcolsep}{4pt}

\begin{tabular}{@{}lcc|c|cc|c|c@{}}
\toprule
& \multicolumn{2}{c|}{\textbf{Frame reinjection}}
& \textbf{Fusion}
& \multicolumn{2}{c|}{\textbf{Representation}}
& \textbf{Readout}
& \textbf{Reconstruction} \\

\cmidrule(lr){2-3}
\cmidrule(lr){4-4}
\cmidrule(lr){5-6}
\cmidrule(lr){7-7}
\cmidrule(l){8-8}

Metric
& On
& Off (ref.)
& Temp. query
& Freq. only
& Temp. only
& Band pooling
& Linear interp. \\

\midrule

F1@0.2 &
\textbf{0.8080}$\pm$\textbf{0.0040} &
0.8006$\pm$0.0051 &
0.5137$\pm$0.0204 &
0.7898$\pm$0.0030 &
0.4024$\pm$0.0125 &
0.6779$\pm$0.0119 &
0.7856$\pm$0.0089 \\

F1@0.3 &
\textbf{0.8075}$\pm$\textbf{0.0042} &
0.7989$\pm$0.0054 &
0.4921$\pm$0.0230 &
0.7892$\pm$0.0044 &
0.3707$\pm$0.0148 &
0.6490$\pm$0.0132 &
0.7828$\pm$0.0097 \\

F1@0.5 &
\textbf{0.7766}$\pm$\textbf{0.0050} &
0.7662$\pm$0.0058 &
0.3909$\pm$0.0298 &
0.7442$\pm$0.0213 &
0.2606$\pm$0.0112 &
0.5479$\pm$0.0142 &
0.7398$\pm$0.0170 \\

\bottomrule
\end{tabular}
\end{table*}

\vspace{-5mm}
\section{Experimental Setup}
\vspace{-2mm}
Experiments were conducted using EEG segments from MASS \cite{oreilly2014mass} and the corresponding MODA expert-consensus spindle annotations \cite{lacourse2020moda}. The analysis included 749 N2 EEG excerpts from 180 subjects, containing 5,342 consensus spindle events. EEG signals were resampled to 128 Hz and analysed as 115-s excerpts.

Fig.~\ref{fig:architecture} depicts the nominal EEG configuration. Inputs were C3, F3 and O1; when F3 was unavailable in
cohort 01-04, upper vertical EOG served as the third downstream channel.
Following SUMO \cite{kaulen2022sumo}, 36 subjects formed a fixed test set and
the remaining 144 were split into 120/24 training/validation subjects across
six folds.

The pretrained SleepFM encoder remained frozen and was employed with the reduced BAS channel configuration described above. Task-specific modules were
trained for $40$ epochs with AdamW (LR $2.66\times10^{-4}$, weight decay
$2.12\times10^{-3}$, batch $16$, gradient clip $1.36$, token stride 192), focal
$(\alpha,\gamma)=(0.707,1.397)$, Smooth L1 $\beta=0.085$, and
$(\lambda_{\mathrm{seg}},\lambda_{\mathrm{bnd}},\lambda_{\mathrm{cls}},
\lambda_{\mathrm{align}})=(1.151,0.964,0.855,0.539)$. LR was reduced by
0.518 after 5 epochs without improvement. Forty Optuna
\cite{akiba2019optuna} trials (seed $42$) on the first validation fold optimized
$\mathrm{F1@0.2}+0.5\,\mathrm{F1@0.5}$; the selected hyperparameters were
fixed across folds. At each epoch, each fold's validation set selected the
probability threshold ($0.40$--$0.70$, step $0.01$), merge gap ($0$--$0.20$ s, step
$0.05$ s), and checkpoint maximizing $F1@0.2$.

At inference, thresholding defined candidate events and probability-weighted
medians aggregated boundaries. One-to-one temporal-IoU matching reported F1
at 0.2, 0.3 and 0.5. Validation-selected settings were applied unchanged to
the fixed test set; SUMO \cite{kaulen2022sumo} and BLAST
\cite{guo2025blast} were retrained and independently optimized under the
same protocol.


\section{Results and Discussion}
\subsection{Detection and Localization}

\begin{figure}[t]
    \centering
    \includegraphics[width=\columnwidth, trim=0cm 0cm 0cm 0cm, clip]{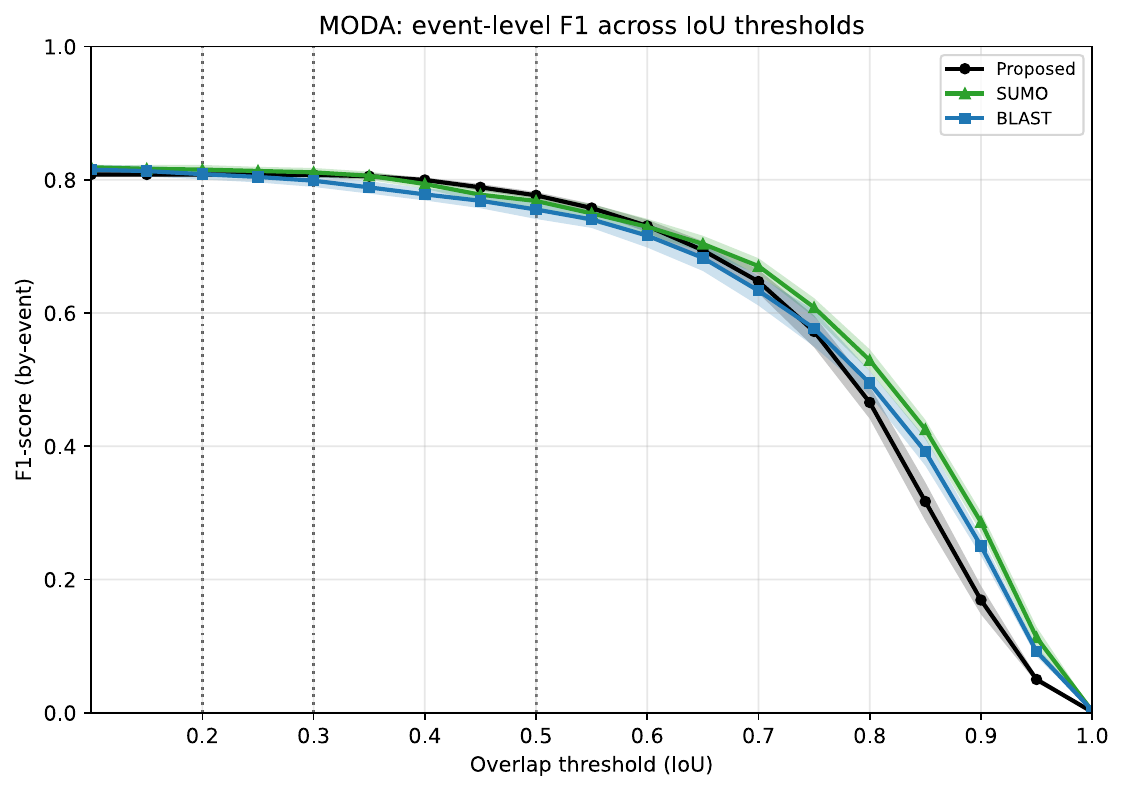}
    \caption{Mean event-level F1 across IoU thresholds over six training/validation folds on the fixed MODA test set.}
    \label{fig:moda_iou}
\end{figure}

Figure~\ref{fig:moda_iou} compares the proposed model, SUMO, and BLAST across IoU thresholds. At IoU $= 0.2$, the proposed model achieves an event-level F1 of 0.8080$\pm$0.0040, remaining comparable to SUMO and BLAST. At IoU $= 0.5$, it reaches 0.7766$\pm$0.0050, compared with 0.7682$\pm$0.0103 for SUMO and 0.7554$\pm$0.0142 for BLAST. At stricter overlap thresholds, however, the proposed model declines more rapidly, with SUMO and subsequently BLAST surpassing it. These results indicate strong spindle-extent recovery at moderate overlap thresholds, while very precise boundary alignment remains challenging.

\subsection{Architectural Contributions}
Table~\ref{tab:ablations} isolates component contributions. With frame reinjection disabled, the fused frequency-query reference achieves F1@0.2 of $0.8006$, versus $0.7898$ for the frequency-only and $0.4024$ for the temporal-only SleepFM. Reversing the cross-attention direction reduces F1 to $0.5137$. These results show that the local time-frequency representation provides the primary localization evidence, while the pretrained SleepFM representation contributes as complementary temporal context.
Band pooling reduces F1 from $0.8006$ to $0.6779$ at IoU $0.2$, and from $0.7662$ to $0.5479$ at IoU $0.5$. Localized wavelet reconstruction outperforms linear interpolation, with the improvement increasing from $0.0150$ at IoU $0.2$ to $0.0264$ at IoU $0.5$. Frame reinjection yields an additional gain, increasing F1 from $0.8006$ to $0.8080$ at IoU $0.2$ and from $0.7662$ to $0.7766$ at IoU $0.5$. These findings suggest that preserving intra-token information and explicitly reconstructing sub-token temporal structure improves boundary-sensitive spindle localization.

\vspace{-3mm}
\section{Conclusion}
This work adapts a pretrained SleepFM model to dense spindle localization
by preserving sub-token information. On MODA, F1 reaches
0.8080$\pm$0.0040 at IoU $0.2$ and 0.7766$\pm$0.0050 at IoU $0.5$,
although stricter thresholds expose residual boundary errors. Ablations
identify local spectral and intra-token information as the main localization
evidence, with SleepFM providing complementary context; structured readout,
localized reconstruction and reinjection further improve recovery. Overall,
dense adaptation benefits from preserving and reconstructing sub-token
information.

\vspace{-3mm}
\section{Compliance with Ethical Standards}

MASS \cite{oreilly2014mass} data were analysed in accordance with the archive’s data-use agreement. The study was conducted under protocol CER-2025-Navarro-PREDSomADICE-00578, which received a favourable opinion from the Research Ethics Committee of Sorbonne Université.
\section{Acknowledgments}
This work was initiated and supported by the DREAM project within the Medical Artificial Intelligence (MedAI) group at Idiap Research Institute funded by Fondation AP-HP. Authors declare no conflicts of interest.
\bibliographystyle{IEEEbib}
\bibliography{references}

\end{document}